\documentclass[twocolumn]{aa}
\usepackage{graphicx,epsfig}

\newcommand{\OmM}{\Omega_{\rm M}}
\newcommand{\OmL}{\Omega_\Lambda}

\newcommand{\thetac}{\theta_{\rm c}}

\newcommand{\fwhm}{\theta_{\rm fwhm}}

\newcommand{\Mdet}{M_{\rm det}}

\newcommand{\yo}{y_{\rm o}}	
\newcommand{\jnu}{j_\nu}		
\newcommand{\Psit}{\Psi_{\theta}}
\newcommand{\Tc}{T_{\theta}}
\newcommand{\sigt}{\sigma_\theta}
\newcommand{\Ft}{F_\theta}
\newcommand{\Pnoise}{P^{\rm noise}}
\newcommand{\Psky}{P^{\rm sky}}
\newcommand{\Pps}{P^{\rm ps}}
\newcommand{\Snu}{S_\nu}
\newcommand{\Slim}{S_{\rm lim}}
\newcommand{\No}{N_{\rm o}}
\newcommand{\So}{S_{\rm o}}
\newcommand{\Bnu}{B_\nu}
\newcommand{\sigpsr}{\sigma_{\rm r}}
\newcommand{\sigpsir}{\sigma_{\rm ir}}
\newcommand{\sigY}{\sigma_{\rm Y}}
\newcommand{\Gr}{G_{\rm r}}
\newcommand{\Gir}{G_{\rm ir}}
\newcommand{\nur}{\nu_{\rm r}}

\newcommand{\nuir}{\nu_{\rm ir}}

\begin{document}
   \title{Point Source Confusion in SZ Cluster Surveys}

   \subtitle{}

   \author{James G. Bartlett\inst{1} \and 
	    Jean--Baptiste Melin\inst{2}  
          }

   \offprints{J. G. Bartlett}

   \institute{\inst{1}APC, 11 pl. Marcelin Berthelot, 75231 
             Paris Cedex 05, FRANCE \\
             (UMR 7164 CNRS, Universit\'e
             Paris 7, CEA, Observatoire de Paris)\\
	      \inst{2}Department of Physics, University of California Davis,
                     One Shields Avenue, Davis, CA , 95616 USA\\
               \email{bartlett@cdf.in2p3.fr, melin@bubba.physics.ucdavis.edu}
             }

   \date{}

   \abstract{We examine the effect of point source confusion on
   cluster detection in Sunyaev--Zel'dovich (SZ) surveys.  A filter
   matched to the spatial and spectral characteristics of the SZ
   signal optimally extracts clusters from the astrophysical
   backgrounds.  We
   calculate the expected confusion (point source and primary cosmic
   microwave background [CMB]) noise through this
   filter and quantify its effect on the detection threshold for both
   single and multiple frequency surveys.  Extrapolating current radio
   counts, we estimate that confusion from sources
   below $\sim 100\;\mu$Jy limits single--frequency surveys to
   $1\sigma$ detection thresholds of $Y\sim 3\times 10^{-6}$~arcmin$^2$ at
   30~GHz and $Y\sim  10^{-5}$~arcmin$^2$ at 15~GHz (for
   unresolved clusters in a 2~arcmin beam);  
   these numbers are highly uncertain, and an
   extrapolation with flatter counts leads to much lower confusion limits.
   Bolometer surveys must contend with an important population of
   infrared point sources. We find that a three--band matched
   filter with 1~arcminute resolution (in each band) 
   efficiently reduces confusion, but does not eliminate it: 
   residual point source and CMB fluctuations contribute significantly
   the total filter noise.  In this light, we find that a 3--band 
   filter with a low--frequency channel (e.g, 90+150+220~GHz) extracts 
   clusters more effectively than one with a high frequency channel (e.g,
   150+220+300~GHz).
 }
\titlerunning{Point Source Confusion in SZ Cluster Surveys}
\authorrunning{Bartlett \& Melin}
   \maketitle
%

\section{Introduction}

Galaxy cluster surveys based on the Sunyaev--Zel'dovich (SZ) effect
(\cite{sun70,sun72}; for comprehensive reviews, see
\cite{birk99,car02}) will soon supply large, homogeneous catalogs out
to redshifts well beyond unity
(\cite{barb96,eke96,col97,hol00,bar01,kne01,ben02}).  Eagerly awaited, these
surveys will probe dark energy and its evolution, and shed new light
on galaxy formation (\cite{hai01,wel03,wang04}).  The instruments
designed for these observations are of two types: dedicated
interferometer arrays surveying at a single frequency (e.g., AMI,
AMiBA, and SZA) and bolometer systems operating over several
millimeter bands (e.g., ACBAR, ACT, APEX, Olimpo, Planck,
SPT)\footnote{A list of experiment web pages is given at the end of
the reference section.}.  Ground--based and balloon--borne instruments
are expected to find up to several thousands of clusters, while the
Planck mission will produce an essentially all--sky catalog of several
$10^4$ clusters by the end of the decade.

An important issue facing these surveys is cluster detection in the
presence of other astrophysical foregrounds/backgrounds.  Except for
the very nearby ones, clusters will appear as extended sources over
arcminute scales.  Power in diffuse Galactic emission (synchrotron,
free--free and dust emission) and in the primary cosmic microwave
background (CMB) anisotropy falls on these scales and the clusters can
be efficiently extracted using an adapted spatial filter
(\cite{hae96,her02,sch04}); fluctuations caused by point
sources (radio and infrared galaxies), on the other hand, are
important on these scales and represent a potentially serious source
of confusion for SZ cluster searches (\cite{knox04,wh04,agh04}).

The two kinds of SZ survey instruments deal with this problem in
different ways.  Single frequency surveys must individually identify
and remove point sources with high angular resolution (better than 1
arcminute) observations, which interferometers achieve by incorporating
several long baselines in their antenna array.  Operating at
relatively low frequencies (15~GHz for AMI, 30~GHz for the SZA and
90~GHz for AMiBA), these surveys will contend with the radio galaxy
population.  Bolometer--based instruments will not have the angular
resolution needed to spatially separate point sources from galaxy
clusters; they must instead rely on spectral information.  In their
millimeter wavelength bands, these instruments will contend with the
poorly known far--infrared point source population.

In this paper we quantify the effect of point source confusion on
cluster detection in SZ surveys.  We shall only consider the effect of
the random field population, but we note that point sources are
expected to preferentially reside in the clusters themselves, locally
raising the effective noise level; we leave examination of this effect
to a future work.  White \& Majumdar (2004) recently calculated the
expected confusion due to IR point sources as a function of frequency,
while Knox et al. (2004) and Aghanim et al. (2004) studied their
influence on the measurement of SZ cluster parameters.  We extend this
work by considering cluster detection with an optimal filter spatially
and spectrally (for multi--frequency surveys) matched to the thermal SZ signal
(\cite{hae96,her02,sch04}).  Using the matched filter, we quantify the
confusion noise induced by extragalactic point sources for both
single--frequency radio and multi--frequency bolometer SZ surveys.

We begin by briefly describing our matched filter and cluster detection
routine, leaving details to Melin et al. (2005).  In Section
\ref{sec:ptsources} we present our point source model, based on recent
number counts in the radio and far--infrared. We then calculate the
confusion noise through the filter to examine its importance for
future SZ surveys (Section~\ref{sec:confnoise}).  In the last section,
we summarize our main results and discuss implications for SZ surveying.

\section{Detecting Clusters: matched SZ filters}

The SZ effect is caused by the hot gas ($T\sim 1-10$~keV) contained in
galaxy clusters known as the intracluster medium (ICM); electrons in
this gas up--scatter CMB photons and create a unique spectral distortion
that is negative at radio wavelengths and positive in the
submillimeter, with a zero--crossing near 220~GHz.  The form of this
distortion is universal (in the non--relativistic limit applicable to
most clusters), while the amplitude is given by the Compton $y$
parameter, an integral of the gas pressure along the line--of--sight.
In a SZ survey, clusters will appear as sources extended over
arcminute scales (apart from the very nearby objects, which are
already known) with brightness profile
\begin{equation}
\Delta i_\nu(\vec{x}) = y(\vec{x}) \jnu
\end{equation}
relative to the mean CMB brightness.
Here $y(\vec{x})$ is the Compton $y$ parameter at position $\vec{x}$ (a
2D vector on the sky) and $\jnu$ is the SZ spectral function evaluated at 
the observation frequency $\nu$.

A SZ survey will produce maps of the sky at one or more
frequencies\footnote{Interferometers actually observe visibilities in
the Fourier plane, although in this paper we will model interferometer
data by its image--plane map.  This should be a reasonable
approximation if sampling in the Fourier plane is sufficiently
good. However, it should be emphasized that the question of which
space is best suited for cluster detection is important and currently
under study.}.  We model the survey data as a vector field
$\vec{M}(\vec{x})$ whose components correspond to these maps (for a
single--frequency survey, the data is a scalar field).  It is a sum of
cluster profiles at positions $\vec{x}_i$ plus noise and foregrounds:
$\vec{M}(\vec{x}) = \sum_{i} y_i(\vec{x}-\vec{x}_i)\vec{\jnu} +
\vec{N}(\vec{x})$, where $\vec{\jnu}$ is the column vector with
components given by $\jnu$ evaluated at the different observation
frequencies.  The vector $\vec{N}(\vec{x})$ includes all non--SZ
signals as well as instrumental noise; we model it as a stationary
random variable with zero mean: $\langle \vec{N} \rangle_N = 0$, where
the average is taken over realizations of both the instrumental noise and
foreground fields.  We thus assume that the mean intensity of the map
is zero, i.e., that the zero mode has been taken out by the
observations\footnote{In principle this affects the measured cluster
$Y$ values, but in practice any effect is small for upcoming
surveys.}.  Although the model of a stationary random variable applies
to the primary CMB anisotropy and (perhaps) the noise, one may
question its suitability for Galactic foregrounds; it does, all the
same, seem a reasonable approximation for fluctuations around the mean
foreground intensity over angular scales pertinent to galaxy cluster
detection (arcminutes).

\begin{figure*}
\includegraphics[scale=1.1]{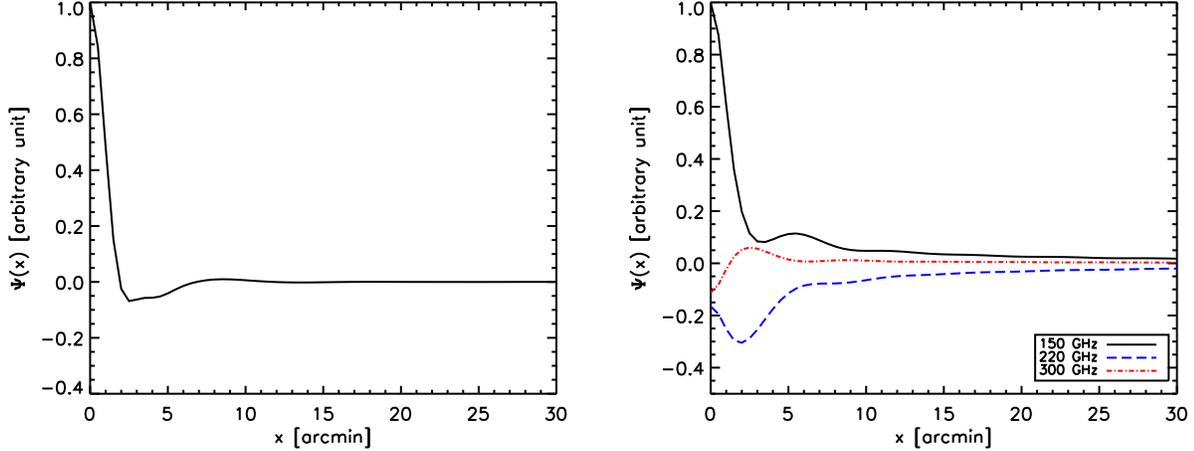}\\
\caption{Radial profiles of single frequency and multiple frequency
matched filters for a cluster of $\theta=1$~arcmin.  In the left--hand 
panel we see the spatial weighting used by the single frequency
filter to optimally extract the cluster from the background 
(radio point sources + CMB) and noise.  The filter is arbitrarily 
normalized to unity at the origin and the beam has a 2~arcmin FWHM.  
The right--hand panel shows the elements of the 3--band filter 
with 150, 220 and 300~GHz channels.  The three functions 
are arbitrarily normalized to the peak value of the 150~GHz 
filter (black curve), and each channel has a 1~arcmin FWHM beam. 
In this case we see how the filter uses both spatial 
and spectral information to optimally extract the cluster.
}
\label{fig:filters}
\end{figure*}

We now wish to use both spatial and spectral information (when
available) to best extract clusters from our survey.  Consider a
cluster of characteristic size $\theta$ (in the following, we take
this to be the core radius of a $\beta$--profile) and central
$y$--value $\yo$ situated at an arbitrary point $\vec{x}_i$ on the
sky.  We build a filter $\vec{\Psit}(\vec{x})$ (in general a column
vector in frequency space) that returns an estimate, $\hat{\yo}$, of
$\yo$ when centered on the cluster:
\begin{equation}\label{eq:filter}
\hat{\yo} = \int d^2x\; \vec{\Psit}^t(\vec{x}-\vec{x}_i) 
      \cdot \vec{M}(\vec{x})
\end{equation}
where superscript $t$ indicates a transpose (with complex conjugation
when necessary).  Suppose that clusters of this characteristic size
are described by an average angular form $\Tc(\vec{x})$, by which we
mean: $\Tc(\vec{x})=\langle y(\vec{x})/\yo\rangle_C$, where the
average is over many clusters of size $\theta$.  We {\em match} the
filter to this angular form $\Tc$ and to the spectral shape of the SZ
signal, requiring an unbiased estimate of the central $y$ value:
$\langle \hat{\yo} \rangle = \yo$, where now the average is over both
total noise and cluster (of size $\theta$) ensembles.  The filter is
then uniquely determined by further demanding a minimum variance estimate.

The result expressed in Fourier space (the flat sky approximation is
reasonable on cluster angular scales) is (\cite{hae96,her02,mel05}):
\begin{equation}
\vec{\Psit}(\vec{k}) = \sigt^2 \vec{P}^{-1}(\vec{k})\cdot \vec{\Ft}(\vec{k})
\end{equation}
where 
\begin{eqnarray}
\vec{\Ft}(\vec{k})   & \equiv & \vec{\jnu} \Tc(\vec{k})\\
\label{eq:sigt}
\sigt          & \equiv & \left[\int d^2k\; 
     \vec{\Ft}^t(\vec{k})\cdot \vec{P}^{-1} \cdot
     \vec{\Ft}(\vec{k})\right]^{-1/2}
\end{eqnarray}
with $\vec{P}(\vec{k})$ being the noise power spectrum, a matrix in
frequency space with components $P_{ij}$ defined by $\langle
N_i(\vec{k})
N_j^*(\vec{k}')\rangle_N=P_{ij}(\vec{k})\delta(\vec{k}-\vec{k}')$.  Note
that this last expression treats, again, the astrophysical foregrounds
as stationary fields.  The quantity $\sigt$ gives the the total noise
variance through the filter.  We write the noise power spectrum as a
sum $P_{ij}=\Pnoise_i\delta_{ij}+B_i(\vec{k})B^*_j(\vec{k})\Psky_{ij}$, where
$\Pnoise_i$ represents the instrumental noise power,
$B(\vec{k})$ the observational beam and $P^{\rm sky}_{ij}$ 
gives the foreground power
(non--SZ signal) between channels $i$ and $j$.  As written, we assume
uncorrelated instrumental noise between observation frequencies.

Our aim is to quantify the effect of point source confusion on cluster
detection using this filter.  To this end, we ignore diffuse Galactic
emission, which is small on cluster scales, and only include primary
CMB temperature anisotropy and point source fluctuations in the sky
power $\Psky_{ij}$.  For our numerical results, we adopt a standard
flat concordance CMB power spectrum ($\OmL=1-\OmM=0.7, h=0.7$, e.g.,
\cite{freed01,spe03}) and employ a cluster template based on the
spherically symmetric $\beta$--model with core radius $\theta$ and
$\beta=2/3$: $\Tc(\vec{x})=(1+|\vec{x}|^2/\theta^2)^{-(3\beta-1)/2}$.

Two examples of the matched filter for $\theta=1$~arcmin 
are given in Fig.~\ref{fig:filters}, one for a single frequency
survey with a 2 arcmin beam (left--hand panel) and the other for 
a 3--band filter with 1 arcmin beams at 150, 220 and 300~GHz (right--hand
panel).  The filters are circularly symmetric because we have chosen a 
spherical cluster model, and the figure shows their radial profiles.
We clearly see the spatial weighting used by the single frequency filter 
to optimally extract the cluster from the noise and from
the point source and CMB backgrounds.  The multiple frequency 
filter $\vec{\Psit}$ is a 3--element column vector containing 
filters for each individual frequency.  Their radial profiles are
shown in the right--hand panel arbitrarily normalized to the peak of 
the 150~GHz filter. The map from each band is filtered by the 
corresponding function and the results are then summed to produce the final
filter output.  We see here how the filter uses both spectral and spatial
weighting to optimally extract the cluster signal.

\section{Point Sources}
\label{sec:ptsources}

\begin{figure}
\includegraphics[scale=0.5]{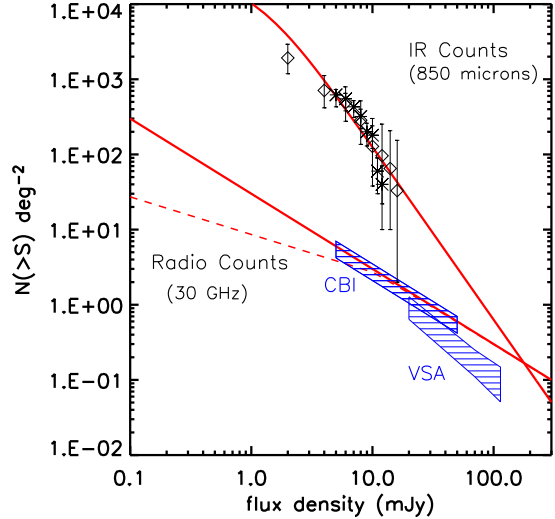}\\
\caption{Integrated radio and IR source counts. 
The lower solid red line results from the differential counts of
Eq.~(\ref{eq:radiocounts}), while the dashed red line corresponds to
the model of Eq.~(\ref{eq:radiocounts_alt}).  Measured counts at
30~GHz from CBI and the VSA are shown as the hashed blue boxes.  The
upper solid red line gives the submillimeter source counts from
Eq.~(\ref{eq:ircounts}).  The diamonds show the measured counts at
$850\;\mu$m from the HDF--North (\cite{bor03}) and the
stars those from the 8mJy--survey (\cite{sco02}).
}
\label{fig:countsfig}
\end{figure}

Two different extragalactic point source populations affect SZ
observations (see Figure~\ref{fig:countsfig}).  At frequencies below
$\sim 100$~GHz, radio galaxies and quasars dominate the source counts,
while at higher frequencies dusty IR galaxies become more important.
The spectral dependence of source flux density in both populations is
often modeled as a power law, $\Snu\propto \nu^\alpha$, with spectral
index $\alpha$.  Radio sources tend to have falling spectra with
$\alpha<0$ (\cite{her92,tay01,mas03}), but flat and inverted spectra
with $\alpha \geq 0$ appear more prominent with observing frequency
(\cite{ben03,tru03}).  At millimeter wavelengths we observe the dust
emission of IR galaxies in the Rayleigh--Jeans with rising spectra
characterized by $\alpha \sim 3-4$ (\cite{vla04}).  As will be seen,
we need information on these extragalactic sources down to mJy
flux densities and below; unfortunately, neither the distribution of
$\alpha$ nor the source counts are well known for either population at
these flux densities and at frequencies of interest for SZ
observations ($\sim 10-300$~GHz).

\subsection{Radio Galaxies}

Bennett et al. (2003) summarize the radio source counts at $30-40$~GHz from
WMAP ($\Snu>1$~Jy), DASI ($\Snu>100$~mJy, \cite{kov02}), VSA
($\Snu>50$~mJy, \cite{tay03,cleary04}) and CBI ($\Snu>10$~mJy,
\cite{mas03}).  Fitting to these data, Knox et al. (2004) find
\begin{equation}\label{eq:radiocounts}
\frac{dN}{d\Snu}\bigg|_{\rm r} = \frac{\No}{\So}
	\left(\frac{\Snu}{\So}\right)^\gamma
\end{equation}
with $\No=30$~deg$^{-2}$, $\So=1$~mJy and $\gamma = -2.0$.  The
integrated counts obtained from Eq.~(\ref{eq:radiocounts}) are shown in
Figure~\ref{fig:countsfig} as the lower solid red line, together with the
observations from CBI and the VSA. The model fits the CBI counts well,
but lies high relative to the VSA counts at the bright end; as 
shown in Bennett et al. (2003), the counts in fact steepen toward higher
flux densities, so a pure power law only matches the data over a
limited range.  

We will be interested in the counts at flux densities near
$100\;\mu$Jy for calculating the expected confusion noise in upcoming
SZ surveys.  This requires an extrapolation of the observed counts
using Eq.~(\ref{eq:radiocounts}) to much lower flux densities, which
we view with caution.  To get a handle on the uncertainty associated
with this extrapolation, we consider an alternate model with a
flattening slope toward the faint end:
\begin{equation}\label{eq:radiocounts_alt}
\frac{dN}{d\Snu}\bigg|_{r'} = \frac{\No}{\So} 
     \left(1 + \frac{\Snu}{\So}\right)^{\gamma}
\end{equation}
with the same values of $\No$, $\So$ and $\gamma$.  These counts are
shown as the red dashed line in Figure~\ref{fig:countsfig}, which
clearly provides an equally satisfactory fit to the observations.
Comparing the confusion noise in the two models will give us a sense
of the uncertainty in our estimates.

We adopt Eq.~(\ref{eq:radiocounts}), alternatively
Eq.~(\ref{eq:radiocounts_alt}), for the counts at $\nu=30$~GHz.
Extrapolation to other frequencies suffers from uncertainty in the
spectral index $\alpha$ of the emission law.  Typically negative,
determinations of $\alpha$ spread over a wide range, including
positive values.  Mason et al. (2003), for example, find an average
$\langle\alpha\rangle=-0.45$, between 1.4 and 30~GHz, with a dispersion
$\sigma_\alpha=0.37$, which is roughly consistent with the
observations of Taylor et al. (2001) between 1.4 and 15~GHz.  The
brighter sources seen over the higher frequency WMAP bands, on the
other hand, show much flatter spectra, with a distribution centered on
$\alpha=0$ and a dispersion of $\sigma_\alpha\sim 0.3$
(\cite{ben03}; see also \cite{tru03}).  There is a pressing need
for better understanding of the radio source population at CMB
frequencies.

\subsection{IR Galaxies}

Dusty infrared luminous galaxies dominate the source counts at
frequencies near 100~GHz and higher. The dust in these galaxies is
typically heated to temperatures of several tens of Kelvin by their
interstellar radiation field.  Its emission can be characterized
with a blackbody spectrum modified by a power--law emissivity: $\Snu
\propto \nu^\beta\Bnu(T)$, where $\beta\sim 1-2$.  In the Rayleigh--Jeans this
leads to a steeply rising power--law with spectral index $\alpha\sim
3-4$, from which we see that source confusion from the IR population
rises rapidly with frequency, the implications of which were recently
discussed by White \& Majumdar (2004).

Blank field counts around 10~mJy at $850\;\mu$m where obtained by
Scott et al. (2002) and Borys et al. (2003) using the SCUBA instrument.  
As discussed by the latter authors, the counts are well described
by a double power--law:
\begin{equation}\label{eq:ircounts}
\frac{dN}{d\Snu}\bigg|_{\rm ir} = \frac{\No}{\So}\left[
	\left(\frac{\Snu}{\So}\right)^{1.0} + \left(\frac{\Snu
	}{\So}\right)^{3.3}\right]^{-1}
\end{equation}
with $\No=1.5\times 10^4$~deg$^{-2}$ and $\So=1.8$~mJy.  This model 
is shown in Figure~\ref{fig:countsfig} as the upper solid red line,
along with data taken from the two surveys.

The SCUBA Local Universe Galaxy Survey (SLUGS, \cite{vla04}) finds a
broad distribution for the dust emissivity index with
$\langle\beta\rangle\sim 1$ and a dispersion we take to be
$\sigma_\beta\sim 0.2$.  According to the SLUGS observations, optical
galaxies tend to have lower spectral indexes than IRAS--selected
objects; we eye--balled the above numbers to be representative of the
population as a whole.

\begin{figure*}
\includegraphics[width=9.5cm,height=9.2cm]{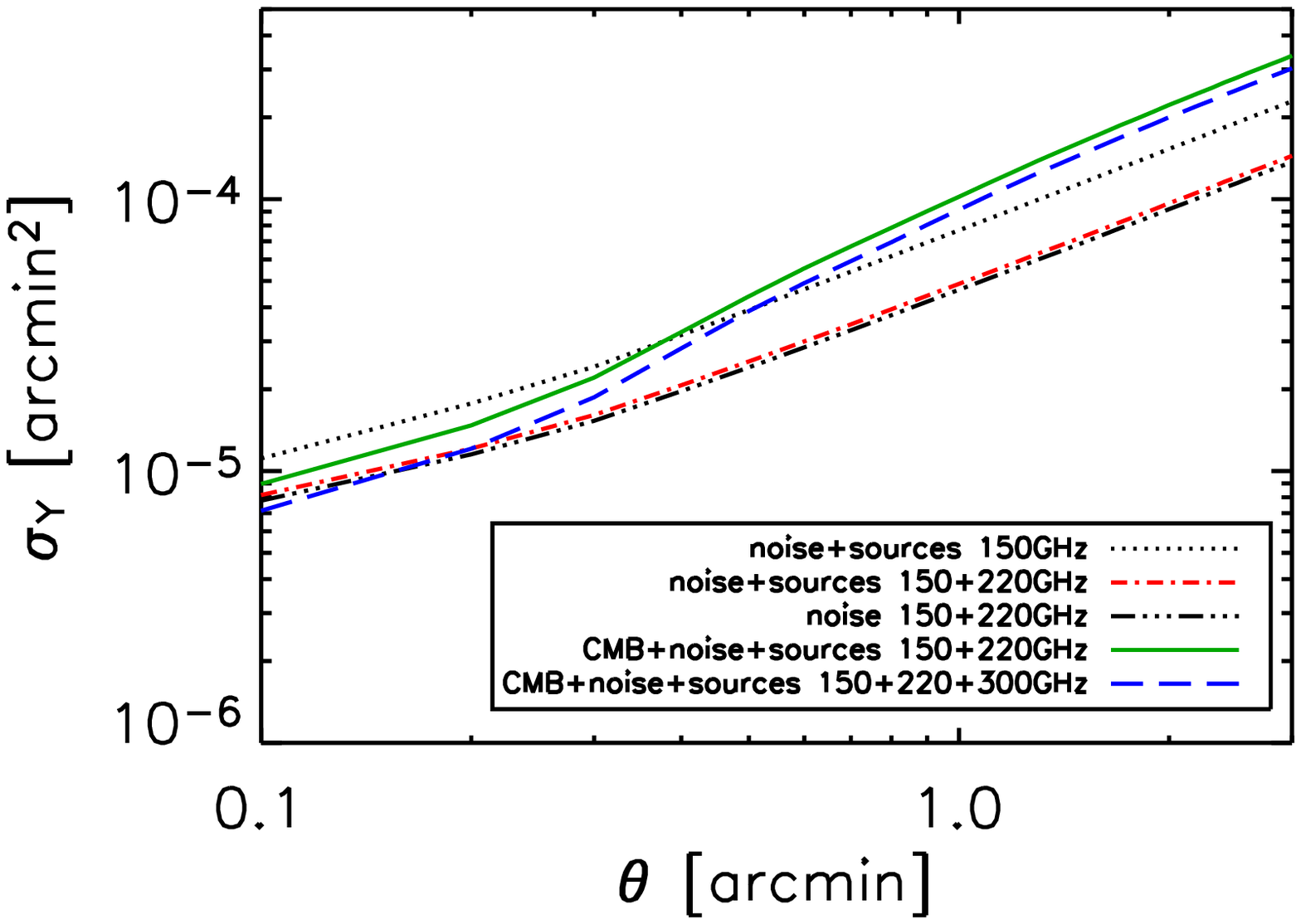}
\includegraphics[width=9.5cm,height=9.2cm]{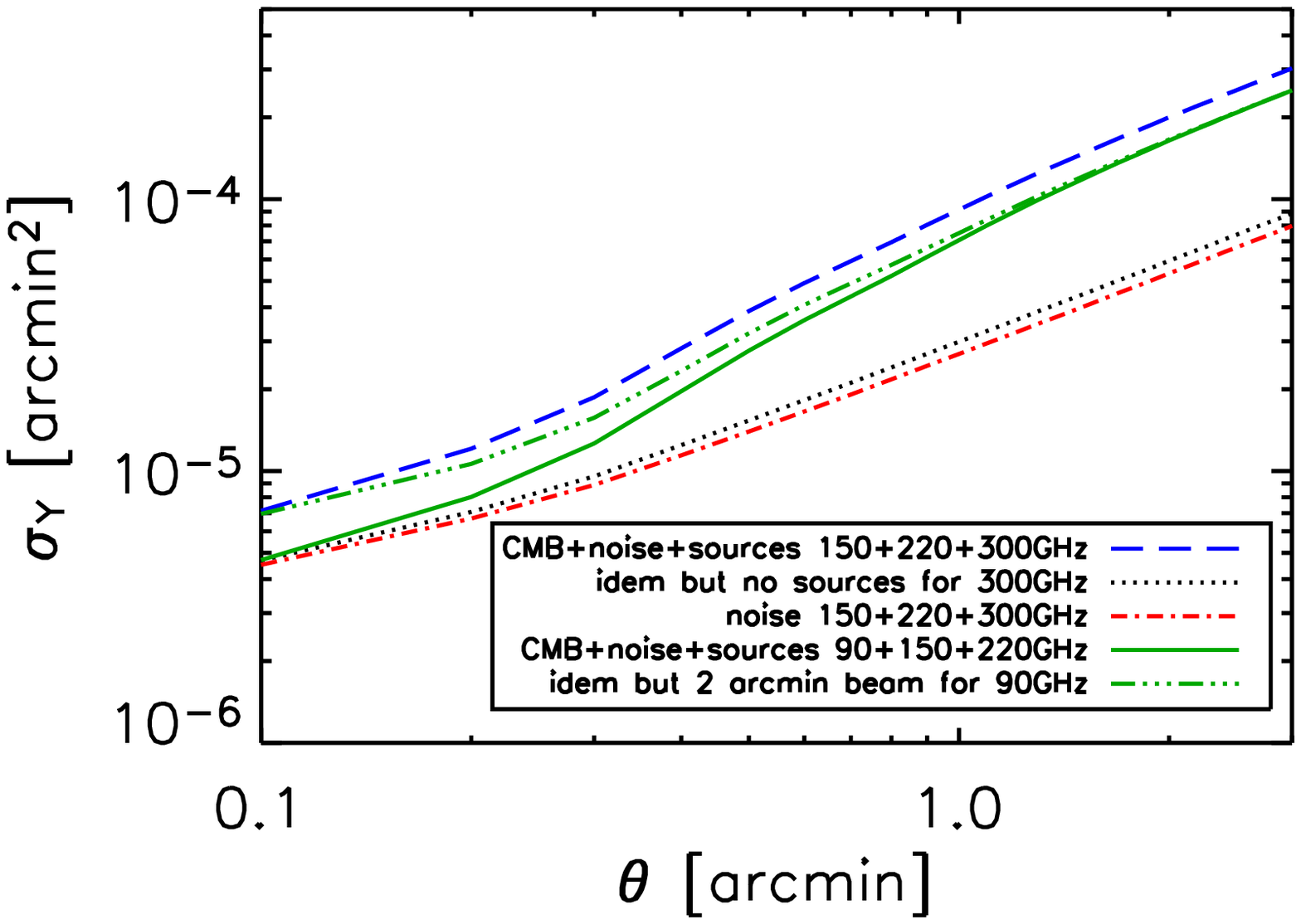}
\caption{Noise ($1\sigma$) on the integrated Compton $Y$ parameter
as a function of filter scale. We assume $10\;\mu$K noise/beam 
and adopt 1~arcmin beams in all bands, except for 
the green 3--dot--dashed curve in the right--hand panel.  At 
these frequencies, we only include IR sources
below $\Slim=100$mJy, following the counts of Eq.~(\ref{eq:ircounts}),
and with a spectral index $\alpha=3$ and no dispersion. {\em Left
Panel:} The black dotted line shows the sensitivity at 150~GHz, where
the instrumental noise and point source confusion are comparable (see
Figure~\ref{fig:spectra}).  Confusion is effectively eliminated by a
2--band filter, as demonstrated by the red dot--dashed line lying just
above the pure instrumental noise limit (black 3 dot--dashed curve).
The upper solid green line shows the result when adding primary CMB
anisotropy, in which case the 2--band filter is unable to cleanly
separate the three astrophysical components.  Addition of a third band
at 300~GHz (blue dashed line) does not greatly improve the result,
because both the 220 and 300~GHz channels are dominated by point
sources, as shown in Figure ~\ref{fig:spectra} (see text).  {\em Right
Panel:} Artificially removing the point sources from the 300~GHz
channel, we see (black dotted line) that the filter gains enough
leverage on the three astrophysical signals to drop its noise to the
instrumental limit (red dot--dashed line).  Source confusion being
greatly reduced at lower frequencies, a 3--band filter with 90~GHz
(solid green line) performs better; this remains true even when 
we degrade the 90~GHz beam to a more realistic 2~arcmin FWHM (green
3--dot--dashed curve).}
\label{fig:sigcy}
\end{figure*}

\section{Point Source Confusion}
\label{sec:confnoise}

Point source confusion is caused by random fluctuations in the number
of unresolved sources in the filter.  We now study the contribution of
this confusion to the overall filter noise, $\sigt$, as a function of
filter scale, $\theta$. As mentioned above, we only consider
uncorrelated instrumental noise, primary CMB anisotropy and point
source terms to the power spectrum matrix $\vec{P}$, whose
off--diagonal elements are then just sky terms (to be multiplied by
the beam):
\begin{equation}
\Psky_{ij}(k) = C_{l=|\vec{k}|} + \Pps_{ij}
\end{equation}
where $C_l$ is the CMB temperature angular power spectrum and
$\Pps_{ij}$ is the point source term.  We quote power in units
of CMB temperature equivalent and ignore spatial correlations of the
point sources, which means that $\Pps_{ij}$ is independent
of $k$.

To calculate the point source terms $\Pps_{ij}$, we adopt the counts
of Eq~(\ref{eq:radiocounts}), alternatively
Eq~(\ref{eq:radiocounts_alt}), at $\nur\equiv30$~GHz and those of
Eq~(\ref{eq:ircounts}) at $\nuir\equiv 350$~GHz ($850\;\mu$m).  Unless
otherwise specified, spectral indexes follow Gaussian distributions
with ($\langle\alpha \rangle=0$, $\sigma_\alpha=0.3$) for radio
sources, and ($\langle\alpha \rangle=3$, $\sigma_\alpha=0.2$) for IR
galaxies (see previous section).  Then we have
\begin{equation}
\Pps_{ij} = \sigpsr^2\cdot R^{\rm r}_{ij}\; \Gr(\sqrt{\nu_i\nu_j}/\nur)
	   + \sigpsir^2\cdot R^{\rm ir}_{ij}\; \Gir(\sqrt{\nu_i\nu_j}/\nuir)
\end{equation}
where (\cite{sch57,con74})
\begin{equation}
\sigma^2_{({\rm r,ir})}  \equiv  
	\left(\frac{\partial B_{(\nur,\nuir)}}{\partial T}\right)^{-2}
	\int d\Snu\; \Snu^2\; 
	\frac{dN}{d\Snu}\bigg|_{({\rm r,ir})} 
\end{equation}
gives the sky temperature variance due to radio or IR sources  
at the fiducial frequencies $\nur$ and $\nuir$; conversion to 
CMB temperature units is made with the Planck function $B_\nu$
at $T_{\rm cmb}=2.725$~K (\cite{mat99}).  This also appears in
the R--factors:
\begin{equation}
R^{({\rm r,ir})}_{ij} \equiv \frac{(\partial B_{\nu_i}/\partial T)^{-1}
	(\partial B_{\nu_j}/\partial T)^{-1}}
	{(\partial B_{\nu_{({\rm r,ir})}}/\partial T)^{-2}}
\end{equation}
The function $G$ accounts for spectral variations:
\begin{equation}
\label{eq:Gfunctions}
G_{\rm ({\rm r,ir})}(x) \equiv \int d\alpha\; N_{({\rm r,ir})}
	(\alpha) \; x^{2\alpha}
\end{equation}
with $N_{({\rm r,ir})}$ being the normal distribution for radio and 
IR spectral indexes, respectively.  These results assume that there
is no correlation between flux density $\Snu$ and spectral index,
$\alpha$, so that the joint distribution is separable.

\subsection{Single Frequency Surveys}

Single frequency surveys can only use spatial information to control
point source confusion.  Operating at $\nu=15$~GHz and $30$~GHz,
respectively, AMI and SZA will contend primarily with the radio source
population; AMiBA, on the other hand, must deal with both radio and IR
sources at $\nu=90$~GHz.  The former two interferometers include long
baselines dedicated to identifying and removing point sources at high
spatial frequency on the sky, where they are cleanly
separated from the more extended cluster SZ emission.  Point source
confusion is then caused by the residual population below the
subtraction threshold, $\Slim$.  

Apart from the beam convolution, point source confusion contributes to
the filter variance $\sigt$ in the same manner as instrumental
noise. 
Using the counts of Eq.~(\ref{eq:radiocounts}) we find for the 
confusion power:
\begin{eqnarray}
\Pps_{30, 30} & = & \sigpsr^2 = (12\; \mu{\rm K\cdot arcmin})^2 
	\left(\frac{\Slim}{100\; \mu{\rm Jy}}\right)\\
\nonumber
\Pps_{15,15} & = & \sigpsr^2\cdot R^{\rm r}_{15,15}\cdot \Gr(1/2)\\
\; & > & (48\; \mu{\rm K\cdot arcmin})^2 
	\left(\frac{\Slim}{100\; \mu{\rm Jy}}\right) 
\end{eqnarray}
where the numerical value in the last line assumes $\Gr(1/2)=1$, while
it most certainly is larger.  These values are comparable to target
sensitivities for interferometer surveys, implying that these
instruments will have to subtract sources down to $\sim 100\;\mu$Jy or
better.  Assuming a $\fwhm \sim 2$~arcmin synthesized beam, we
calculate the confusion limit on the integrated Compton $Y$ parameter
for unresolved clusters.  We conclude that confusion from sources
below $100\;\mu$Jy will limit SZ sensitivity to $Y\sim 3\times
10^{-6}$~arcmin$^2$ at $30$~GHz and $Y\sim 10^{-5}$~arcmin$^2$ at
$15$~GHz (both are at $1\sigma$).  The former limit is a factor
$\sim 5$ greater than the confusion expected from primary CMB 
anisotropy in the concordence model for unresolved clusters 
(\cite{mel_sf}).

Adopting the alternative model of Eq.~(\ref{eq:radiocounts_alt}),
these confusion limits drop by a factor $\sim 20$, and the dependence
of the variance on the source subtraction threshold approaches
$\propto\Slim^3$.  There is clearly a large uncertainty associated
with extrapolation of the counts to faint flux densities. In this
light, note that as long as the counts do not steepen toward lower
flux densities, the confusion estimates are, fortunately, dominated by
the counts at $\Slim$, rather than at some unknown cut--off at even
fainter levels.

\begin{figure}
\includegraphics[width=9cm,height=8.8cm]{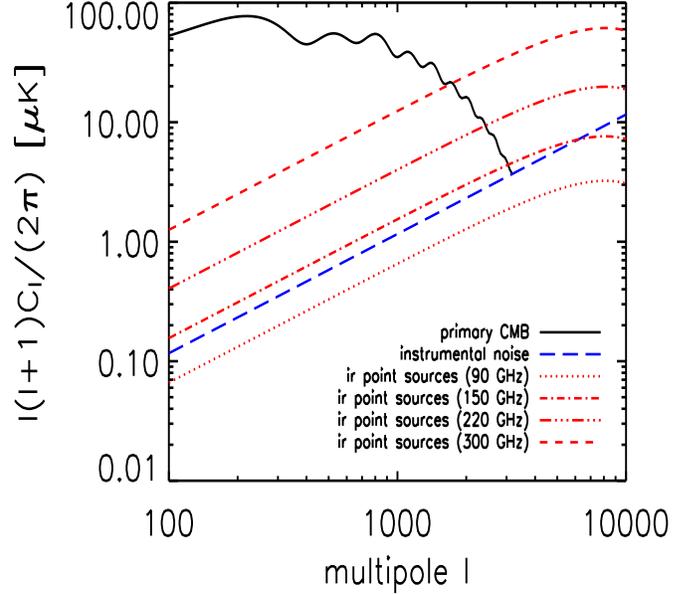}
\caption{Power spectra of the primary CMB anisotropy (solid black line), 
point source confusion in different bands (as labeled) and
instrumental noise; the latter corresponds to $10\;\mu$K
noise/$1$~arcmin lobe, which we take here to be the same in all bands.
Only IR sources with $\Snu<\Slim=100$~mJy are included, according to
the counts of Eq.~(\ref{eq:ircounts}), and with $\alpha=3$ and no
dispersion.  The figure shows the that the 300~GHz channel is
dominated by sources, explaining why the 3--band filter with 90~GHz
performs more efficiently than the one with 300~GHz (see
Figure~\ref{fig:sigcy}).  }
\label{fig:spectra}
\end{figure}

\subsection{Multiple Frequency Surveys}

Planned bolometer--based surveys will operate over several millimeter
and submillimeter bands, allowing them to use spectral information to
extract clusters from the foregrounds.  This will in fact be their
only way to reduce the effects of point source confusion, because they
will not have the spatial resolution\footnote{Most planned surveys
have arcminute resolution.} needed for subtracting point sources from
cluster images.  In the multi--frequency case, we refer to the
optimal spatio--spectral filter as a multi--filter.

Specifically, the multi--filter performs a weighted sum designed to
remove foregrounds from the SZ cluster signal, as illustrated in the
right--hand panel of Figure~\ref{fig:filters}.  Figure~\ref{fig:sigcy}
helps to understand the filter's workings and it will allow us to draw some
important conclusions.  The Figure shows the filter noise $\sigY$
($1\sigma$) in terms of the integrated Compton $Y$ parameter as a
function of filter scale for various band and foreground
combinations.  This is calculated from the filter variance as $\sigY =
\sigt\int d^2x\;\Tc(\vec{x})$.  We take as representative of planned
observations a survey with $10\;\mu$K instrumental noise per 1~arcmin
lobe (FWHM) in all bands.  We only include the IR source population
(dominant at these frequencies) below $\Slim=100$~mJy, assuming
brighter sources are explicitly subtracted; this is well above the
knee in the differential counts of Eq.~(\ref{eq:ircounts}).  We
further fix, for the present, the spectral index to $\alpha=3$ with
zero dispersion ($G=1$).

Consider first the case with just instrumental noise, point sources
(no CMB) and two frequencies, one at $\nu=150$~GHz and the other at the
thermal SZ null, $\nu=220$~GHz.  Point source confusion is severe in
each individual band, as illustrated by Figure~\ref{fig:spectra}.  The
dotted line in the left panel of Figure~\ref{fig:sigcy} shows the
total filter noise with just the 150~GHz band. When both bands are
used in the filter, the filter noise $\sigY$ drops considerably (red
dot--dashed line), approaching the pure instrumental noise limit for
the 2--band filter, shown as the black dashed--3 dotted line.  It is
straightforward to show under these circumstances that the filter
performs a direct subtraction of the point source signal from the
150~GHz channel by extrapolation of the 220~GHz signal using the
known spectral index $\alpha$.  What we see here is that the
subtraction is perfect to the instrumental noise limit.

The sky of course also includes the CMB signal and other foregrounds,
and the point source spectral index has a non--zero distribution,
both of which complicate the subtraction.  We discuss the second
effect below and now add CMB anisotropy, keeping $\alpha$ fixed.  The
2--band filter is no longer able to separately determine the three sky
signals (SZ, point sources and CMB), with as a consequence a
significant rise in total SZ noise, as shown by the solid green line.
Surprisingly, the situation does not improve even when we include more
information with a third observing band at $\nu=300$~GHz (blue dashed
line).  

This interesting result is due to the fact that both the 220 and
300~GHz bands are dominated by point source confusion (see
Figure~\ref{fig:spectra}) -- neither provides good information the CMB
anisotropy, so the filter remains unable to completely separate the three sky
signals.  We can test this conclusion by artificially removing point
sources from the new channel at 300~GHz.  The result is shown in the
right--hand panel of the Figure as the black dotted line; the total SZ
noise $\sigY$ drops to a level comparable to the level induced just by
instrumental noise (red dot--dashed line), indicating that once again
the subtraction is almost perfect.

This has important consequences for SZ surveying.  Observing at high
frequencies, such as $300$~GHz, is very difficult from the ground due
to atmospheric effects; moving up in frequency, one approaches strong
water vapor lines.  What we have just seen suggests that including
bands beyond the thermal SZ null may not be worth the effort, at least
not for detecting clusters.  

To further explore this issue, we replace the 300~GHz channel by a
90~GHz band.  At this lower frequency, point source confusion is
greatly reduced and gives the filter a better handle on CMB anisotropy;
in fact, as shown in Figure~\ref{fig:spectra}, point source confusion
is well below the instrumental noise level at 90~GHz.  The green solid
line in the right--hand panel of Figure~\ref{fig:sigcy} shows the new
result: this three--band filter with 90~GHz performs significantly
better than the one with 300~GHz.  It is, however, unlikely that
all three bands will have the same beam size, which we have taken
as 1~arcmin throughout this discussion.  The green 3--dot--dashed 
curve in the right--hand panel of the Figure shows the result for a 
three--band filter with a 2~arcmin beam at 90~GHz.  Even with the larger 
beam at the lower frequency, the result remains qualitatively the same 
-- the filter with 90~GHz performs better than the one with 300~GHz.  

Multi--frequency surveys will
certainly include 150 and 220~GHz bands to cover the maximum decrement
and null of the thermal SZ signal.  We conclude here that a 90~GHz
band is a more valuable addition than one at 300~GHz for cluster
extraction, despite a loss in angular resolution at the lower frequency.

As a final note, we consider the effects of dispersion in the source
spectral index $\alpha$.  With dispersion, the filter can no longer
perform a perfect subtraction by extrapolation across bands; it must
instead find appropriate frequency weights for a statistically optimal
subtraction.  In the point source power spectrum matrix $\Pps_{ij}$,
only the self power $\Pps_{\nuir,\nuir}$ remains unaffected; other
diagonal elements will increase, while correlations between bands
(in the off--diagonal elements) decrease.  We therefore expect the filter's
performance to decline.  

We examined the importance of this effect using the function $\Gir$
defined in Eq.~(\ref{eq:Gfunctions}).  Dispersions up to
$\sigma_\alpha=0.5$ increase matrix elements involving the 150 and
220~GHz bands of the power spectrum matrix by at most 20\%, relative
to their values with zero dispersion.  The 90~GHz channel is of course
the most affected: the auto--power element $\Pps_{90,90}$ increases by
more than 50\% for the same dispersion.  Nevertheless, when running
the filter combinations of Figure~\ref{fig:sigcy}, we find only a
small change in the filter noise curves, barely perceptible by eye.
We conclude that even rather large variations in the frequency
dependence of individual source spectra does not significantly
increase confusion noise through the filter.  In more general terms,
this also suggests that our confusion estimates are not strongly
dependent on the foreground model.

\section{Discussion and Conclusion}

Primary CMB anisotropy and extragalactic point sources are the most
important foregrounds for SZ surveys.  Point source confusion is a
particularly critical issue because it rises rapidly on cluster
angular scales.  We have quantified its importance for both single
frequency and multiple frequency surveys using current estimates of
the radio and IR source counts and an optimal matched multi--filter
for cluster extraction.  Our main conclusion are:
\begin{itemize}
\item The expected confusion level from radio point sources 
is uncertain due to lack of information on the counts at required flux
densities: a power--law extrapolation of the observed counts
(Eq.~\ref{eq:radiocounts}) leads to a confusion--limited SZ
sensitivity at $30$~GHz of \mbox{$Y\sim 3\times 10^{-6}$~arcmin$^2$} from
unresolved sources below $\Slim=100\;\mu$Jy ($1\sigma$ for clusters
unresolved by a 2 arcmin beam); at 15~GHz the corresponding
sensitivity is $Y\sim 10^{-5}$~arcmin$^2$.  An alternative model in
which the counts flatten at low flux density
(Eq.~\ref{eq:radiocounts_alt}) predicts much lower confusion limits,
reduced by a factor of $\sim 20$, in which case CMB confusion becomes
the limiting factor (\cite{mel_sf}).
\item Currently planned multiple frequency bolometer surveys (e.g.,
10$~\mu$K instrumental noise/ 1~arcmin beam) greatly reduce the
large confusion noise from IR sources and can attain 1~$\sigma$ sensitivity
of $Y \sim 7\times 10^{-6}$~arcmin$^2$ for unresolved clusters 
(see Figure~\ref{fig:sigcy}). Nevertheless, both residual 
point source and CMB anisotropy confusion significantly increase the 
total filter noise, degrading survey sensitivity from
the ultimate instrumental noise limit.
\item For multiple frequency surveys, observation bands below the 
thermal SZ null (e.g., 90+150+220~GHz) are more efficient for cluster
extraction than observations at higher frequencies (e.g.,
150+220+300~GHz), despite a potential loss of angular resolution in
the 90~GHz channel (e.g., 2~arcmin beam instead of our fiducial 1~arcmin 
beam).  
\end{itemize}

\begin{figure}
\includegraphics[width=9cm,height=8.8cm]{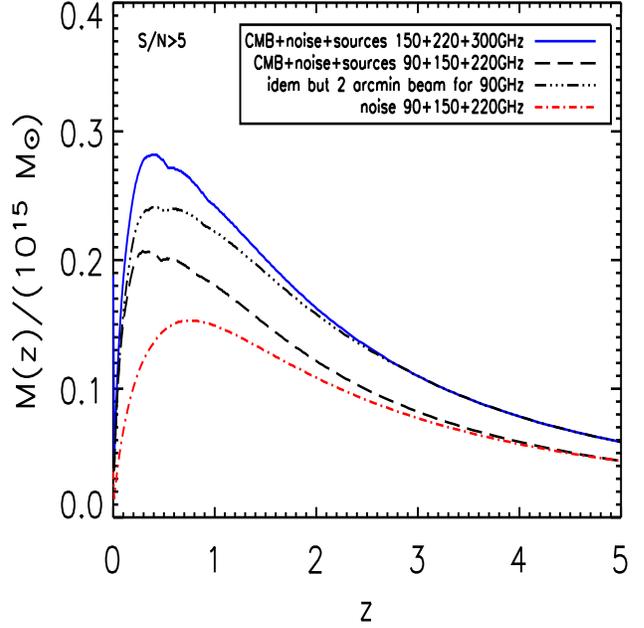}
\caption{Minimum detectable mass as a function of redshift. 
The upper solid blue curve shows the result for the 3--band filter
with a 300~GHz channel, while the middle dashed black curve gives the
result when this channel is replaced by a 90~GHz band.  For reference,
the lower red dash--dotted curve gives the ideal detector--noise
limited detection mass for the 3--band filter with 90~GHz.  The
3--band filter with 90~GHz shows its higher sensitivity (see
Figure~\ref{fig:sigcy}), but still suffers from residual foreground
(CMB) contamination. These results follow for 1~arcmin FWHM beams in
all bands.  For comparison, the black 3--dot--dashed line gives the mass
limit for the 3--band filter with a 90GHz beam of 2~arcmin FWHM, which
continues to perform better than the 3--band filter with 300~GHz despite
the loss of angular resolution at the lower frequency.}
\label{fig:masslimit}
\end{figure}

The minimum detectable cluster mass as a function of redshift,
$\Mdet(z)$, is a key characteristic of a survey.  To further
illustrate the last point, we show $\Mdet(z)$ in
Figure~\ref{fig:masslimit} for the 3--band multi--filter.  In making
this figure, we adopted simple self--similar relations for $Y(M,z)$
and $\thetac(M,z)$ (\cite{mel_sf}); given the potentially important
theoretical uncertainty in these relations, the absolute positioning
of the curves should be viewed with caution -- more robust and more
pertinent to our discussion are their relative positions.  The 3--band
filter with 90~GHz (black dashed line) gains mass sensitivity compared
to the filter with 300~GHz (solid blue line), assuming 1~arcmin beams in 
all bands.  It does not, however, reach the ideal noise limit 
(red dot--dashed line), due to residual point source and CMB confusion 
through the filter.  A 3--band filter with lower angular resolution at
90~GHz (2~arcmin beam), a more probable observing situation, does 
somewhat worse, but still better than the 3--band filter with 300~GHz
(and 1~arcmin beams in all bands). This is an important consideration
given the difficulty imposed by atmospheric effects on observations 
at high frequencies.


With the $S/N>5$ detection threshold [i.e., $5\sigY(\theta)$], we find
15~clusters/deg$^2$ for the filter with 300~GHz (1~arcmin all bands), 
30~clusters/deg$^2$ for the filter with 90~GHz and 1~arcmin beams,
and 18~clusters/deg$^2$ for the 3--band filter with a 2~arcmin 
beam at 90~GHz.
There are 47~detected clusters/deg$^2$ at the ultimate noise--limit 
of the 3--band filter with a 1~arcmin beam at 90~GHz.  These numbers 
should be compared to the
85~clusters/deg$^2$ with mass $>10^{14}$ solar masses with our model
parameters.  

An important consequence of our results is that SZ survey selection
functions are affected by residual astrophysical confusion and are not 
uniquely determined by instrumental properties.  Specifically, we have 
seen that even a 3--band bolometer survey with good angular resolution
and optimal filter cluster extraction experiences a mixture of residual 
point source and primary CMB confusion.  Cluster catalog 
construction will therefore suffer from uncertaintly in astrophysical
foreground modeling (\cite{mel_sf}).

In conclusion, our results support the expectation that future
ground--based SZ surveys will provide rich cluster catalogs for
cosmological studies.  Confusion from point sources and primary CMB
anisotropy can be greatly reduced by multi--frequency
bolometer surveys, but some residual point source and CMB anistropy 
confusion noise will affect cluster detection and catalog construction.

\begin{acknowledgements}
We thank G. Evrard and the organizers of the Future of Cosmology 
with Clusters of Galaxies conference, where this work began in earnest.  
Thanks to J. Mohr for encouragement and to the anonymous referee who 
helped improve the clarity of the presentation. J.-B. Melin was
supported at UC Davis by the National Science Foundation
under Grant N0. 0307961 and NASA under grant No. NAG5-11098.
\end{acknowledgements}

\noindent Web pages of various SZ experiments:
\begin{itemize}{\small
\item ACBAR {\tt http://cosmology.berkeley.edu/group/swlh/acbar/}
\item ACT {\tt http://www.hep.upenn.edu/$\sim$angelica/act/act.html}
\item AMI {\tt http://www.mrao.cam.ac.uk/telescopes/ami/index.html}
\item AMiBA {\tt http://www.asiaa.sinica.edu.tw/amiba}
\item APEX {\tt http://bolo.berkeley.edu/apexsz}
\item SPT {\tt http://astro.uchicago.edu/spt/}
\item SZA {\tt http://astro.uchicago.edu/sze}
\item Olimpo {\tt http://oberon.roma1.infn.it/}
\item Planck {\tt http://astro.estec.esa.nl/Planck/}}
\end{itemize}

\end{document}